\documentclass[12pt]{iopart}
\usepackage{graphicx}

\begin{document}

\title[Encoding information into precipitation structures]{Encoding information into precipitation structures}
%\footnote[1]{presented at the SigmaPhi2008 conference}}

\author{Kirsten Martens$^1$, Ioana Bena$^1$, Michel Droz$^1$ and Zoltan R\'acz$^2$}

\address{$^1$Theoretical Physics Department,
University of Geneva, 1211 Geneva 4, Switzerland}
\address{$^2$Institute for Theoretical Physics--HAS,
E\"otv\"os University, 1117 Budapest, Hungary}

\ead{kirsten.martens@unige.ch, ioana.bena@unige.ch, michel.droz@unige.ch and racz@general.elte.hu}

\begin{abstract}
Material design at submicron scales would be
profoundly affected if the formation of precipitation patterns
could be easily controlled. It
would allow the direct building of bulk structures,
in contrast to traditional techniques which
consist of removing material in order to create patterns.
Here, we discuss an extension of our recent
proposal of using electrical currents to control precipitation
bands which emerge in the wake of reaction fronts in
${\rm A^{+}+B^{-}\to C}$ reaction-diffusion processes.
Our main result, based on simulating the
reaction-diffusion-precipitation equations, is that the dynamics
of the charged agents can be guided
by an appropriately designed time-dependent electric current
so that, in addition to the control of the band spacing, the
width of the precipitation bands can also be tuned.
This makes straightforward the encoding of information into
precipitation patterns and, as an amusing example, we
demonstrate the feasibility by showing how to encode a
musical rhythm.

\end{abstract}

\vspace{10pt}
\begin{indented}
\item{\it Keywords\/}: Pattern formation (Theory), Chemical kinetics, Coarsening processes, Nonlinear dynamics, Kinetic growth processes (Theory)
\end{indented}


\maketitle

%\tableofcontents

\section{Introduction}
\label{Intro}
Information encoding and material design involves the creation of
patterns which, in practice, often means that structures must be
produced in a homogeneous media. At submicron range which is
the target for downscaling of electronic devices, the control
over the desired patterns becomes difficult and, furthermore,
the expenses of traditional {\it top-down} methods (such as e.g.
lithography where material is removed in order to create
structures) grow steeply. A possible way out of
the difficulties is the so called {\it bottom-up} design where
one aims at forming structures directly in the bulk.
Nature, of course, provides us with illuminating
examples of three-dimensional pattern formation at all scales
\cite{{shinbrot},{CrossHoh}}. Among them, there is a much studied
class of reaction-diffusion processes yielding precipitation patterns
\cite{Henisch}, and these processes -- suitably planned and
controlled -- are promising
candidates for {\it bottom-up} designs \cite{lu,grzybowski}. Indeed,
there has been a series of attempts to control the emerging patterns
by appropriately chosen geometry \cite{giraldo} and boundary conditions \cite{grzybowski}, or by a combined tuning of the initial and boundary
conditions \cite{tsapatsis,Guiding-field}. Unfortunately, the above
methods of control are not practical enough,
and more flexible approaches are required.

Recently, we introduced a novel method of pattern control
\cite{Pattern-Design} based on the use of electric currents
for regulating the dynamics of the reaction zones. We showed
both theoretically and experimentally that, by
controlling the reaction zones, the positions of precipitation
bands were predesignable. Here we further develop the theory
of the new method and show that, in addition to the control of the
spacings of the precipitation bands, it is possible to control
the widths of the bands, as well. Thus an extra degree of freedom
appears which can be used to encode information.
We demonstrate the utility of this extra freedom by
encoding rhythmic patterns into precipitation structures.

In order to describe the new features of the control
by electric currents, we begin by a brief description of the
properties of the all important reaction zones which provide
the main input to the precipitation process in the wake of the
zone (Sec.\ref{front}). Next, the effect of
time dependent electric currents is
described and the mathematical details needed for the simulations
of the process are explained (Sec.\ref{curr-front}). Finally
the idea of how to control the width of the
precipitation band is introduced and examples of information
encoded into the widths are presented (Sec.\ref{width}).

\section{Understanding natural precipitation patterns}
\label{front}

The basic idea for the pattern control comes from the observation
that precipitation patterns are often formed in the wake of
moving reaction fronts \cite{CrossHoh,Henisch}. The motion of
the front and its reaction dynamics determines where and when
the concentration of reaction product crosses a threshold
thus inducing precipitation. Consequently, and this is the essence
of our proposal, control over the precipitation pattern can be
realized by regulating the properties of the reaction fronts.
Guiding reaction fronts and tuning the reaction rates
in them, however, does not appear to be an easy task. In order
to explain how it can be done, we turn to the concrete example
of Liesegang patterns \cite{Henisch, Liese-1896}. They have been
studied for more than a century and a wealth of information has
been collected about the properties of the front dynamics
underlying this pattern formation.

\begin{figure}[htbp]
\centering\includegraphics[width=16cm, clip]{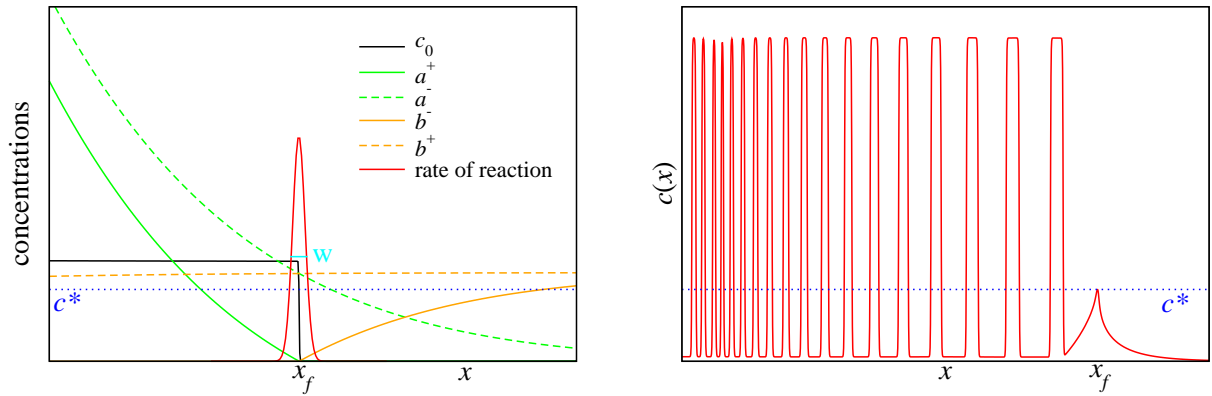}
\caption{Left panel: Concentration profiles near the reaction
zone. Concentrations of the reacting ions ($a^+,b^-$), of the counter ions ($a^-,b^+$), and of the reaction product $c_0$ left in the wake of the front are plotted.
The reaction zone defined as the region where
the rate of $C$ production is significantly different from
zero moves diffusively ($x_f\sim \sqrt{t}$) and remains well
localized during the process.
Right panel: Moment of the creation of a band in the phase separation
process -- the concentration $c$ reaches the threshold
value $c^*$ at the position of the front, $x_f$.} 
\label{fig-front-and-band}
\end{figure}
%%%%%%%%%%%%%%%%%%%%%%%%%%%%%%%%%%%%%%%%%%%%%%%%%%%%%%%%%%
%%%%%%%%%%%%%%%%%%%%%%%%%%%%%%%%%%%%%%%%%%%%%%%%%%%%%%%%%%
\subsection{Properties of a diffusive reaction front}
\label{sec-prop}
Liesegang patterns are characteristic examples of precipitation
structures formed in the wake of moving reaction
fronts \cite{Dee1986}.
The main ingredients are
two electrolytes $A\equiv(A^+,A^-)$ and $B\equiv(B^+,B^-)$
which react with reaction rate $k$ according to the reaction scheme
$A^++B^-\stackrel{k}{\rightarrow} C$. The reaction product $C$
may participate in further reactions but, in the simplest case
considered here, it just undergoes a phase separation
process resulting in an insoluble precipitate provided
the local concentration is above some threshold \cite{ModelB}.
In a typical experiment, the electrolytes are
initially separated with the inner electrolyte $B$ homogeneously
dissolved in a gel column while the outer electrolyte $A$ is
kept in an aqueous solution. At time $t=0$, the outer
electrolyte brought into contact with the end of the gel column
and, since the
initial concentration, $a_0$, of $A$ is chosen to be much
higher than that of $B$ (typically $a_0/b_0\approx 100$),
$A$ invades the gel and a reaction front emerges which advances
along the
column. The motion of this front and the amount
of reaction product $C$ left behind the front are
clearly important factors since they determine the
input for the precipitation processes.

The main features of the reaction front (see left panel of Fig.~\ref{fig-front-and-band}) are well known
\cite{GR1988,MatPack98}
and can be summarized in the following three points:
\begin{description}
\item(i) \hspace{2pt}
The front moves diffusively. Its position $x_f$ is given by
$x_f(t)=\sqrt{2D_ft}$
where the diffusion coefficient $D_f$ is determined
by the initial concentrations ($a_0$, $b_0$) and by the
diffusion coefficients of the reagents.

\item(ii) The front is localized. Although the width $\hbox{w}$ of the front
is slowly increasing with time $(\hbox{w}\sim t^{1/6})$, it is always much
smaller than the diffusive lengthscales $(\sim t^{1/2})$ present
in the problem. Furthermore, the width becomes negligible for fast
reactions $(\hbox{w}\sim 1/k^{1/3})$.

\item(iii)\hspace{-2pt}
The concentration $c_0$ of the reaction product, $C$, left in the wake
of the front is constant with a value depending on the
initial concentrations and on the diffusion constants of the reagents.
\end{description}

Clearly, all these properties are important: (i) tells the
location of the front, (ii) ensures that the position of the
front $x_f(t)$ is precisely given, and (iii) provides
the amount of $C$ produced at the position specified in (i).

\subsection{Phase separation in the wake of the reaction front}

Once the spatial production of $C$-s is specified by (i)-(iii),
the next step of the pattern formation is the phase separation
of the $C$-s. It takes place only if their local concentration $c$
is above some precipitation threshold, $c>c^*$.
The precipitation pattern itself is then the result of a
complex interplay between the production of $C$-s
and the ensuing phase separation dynamics in the wake of
the front. Namely, the experimental parameters ($a_0$ and $b_0$)
are chosen such that $c_0>c^*$ and, consequently, the front
produces a precipitation band at the very beginning. This band
attracts the newly produced $C$-s from the nearby, diffusively
advancing front, thus the concentration $c$ in the front
decreases below $c^*$. As the front moves far enough, the
depletion effect of the band diminishes and the $c>c^*$
condition is satisfied again in the front,
thus leading to the formation of a new band. A quasiperiodic
reiteration of the above process yields the Liesegang
patterns (see right panel of Fig.~\ref{fig-front-and-band}). 
Depending on the details of the phase separation
dynamics, the position of the $n$-th band $x_n$ may vary,
 but there are three well established laws
 that govern the structure of the Liesegang bands:
\begin{description}
\item(a) Time law \cite{timelaw}: The position of the $n$-th
band $x_n$ (measured from the initial interface of the reagents)
is given by $x_n=\sqrt{2D_f t_n}$, where $t_n$
is the time of creation of the band.

\item(b) Spacing law \cite{jabli}: The positions of the bands
form a geometric series $x_n\sim(1+p)^n$ with a spacing
coefficient $p>0$, such that distances between successive
bands increase with the band index $n$.

\item(c) Width law \cite{widthlaw}: The width of the $n$-th
band $w_n$ is proportional to its position: $w_n\sim x_n$.
\end{description}

The above laws can be derived \cite{ModelB}
by using the Cahn-Hilliard equation \cite{cahn58}
for describing of the phase-separation process.
This approach also allows to demonstrate that
the band positions can be controlled
by $a_0$ or $b_0$ since $p$ in the spacing law depends
on these quantities ($p\sim 1/a_0$ is the so called Matalon-Pacter
law \cite{Matalon1955}). Unfortunately, the possible changes
are rather limited since the band positions invariably form a geometric
series.

\section{Description of the control tool}
\label{curr-front}
\subsection{Main idea - controlling the motion of the reagents by electric current}
Equipped with the understanding of both the front
motion and the precipitation processes, we can start to
think of possible control
mechanisms. There are basically two ways to change the structure
of the pattern characterized by the spacing law (b) and the
width law (c). First, one can try to change the functional
form of the time law (a). This can be done by using
various geometries or patterns in the initial state
\cite{grzybowski,giraldo,tsapatsis,inhom},
by employing guiding temperature- or pH fields \cite{Guiding-field},
or by considering systems where the diffusion of the
reacting species is anomalous \cite{Lindenberg2006,deWit2008}.
Unfortunately, these methods are rather unwieldy
and are not flexible enough to easily create arbitrary patterns.

\begin{figure}[htbp]
\begin{center}
\includegraphics[width=8cm, clip]{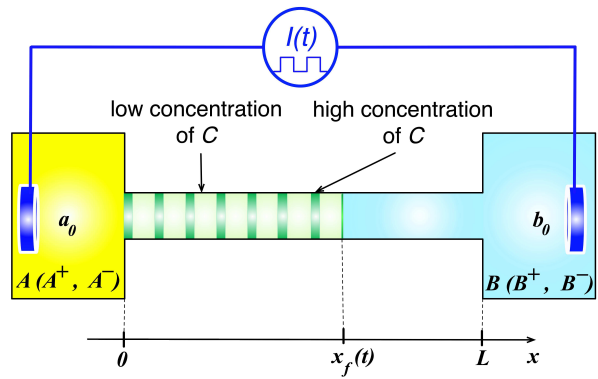}
\caption{Experimental setup for producing Liesegang precipitation patterns
as described in Sec.~\ref{sec-prop}.
The controlling agent is the generator providing electric current
$I(t)$ with a prescribed time-dependence (figure is taken from \cite{Pattern-Design}).}
\label{fig-scheme}
\end{center}
\end{figure}

The second method keeps the time law unchanged
and aims at controlling the creation time $t_n$ of the $n$-th band.
Recalling that $t_n$ is the instant when the concentration
of $C$-s crosses the threshold value $c^*$, one recognizes that
$t_n$ can be controlled by regulating the concentration
$c_0(x)$ at the front. Recently, it was shown both theoretically and
experimentally that the above method can be made to work
by sending a time-dependent electric current through the system
\cite{Pattern-Design}.
The schematic setup for control is shown in
Fig.\ref{fig-scheme} and, at a phenomenological level,
its working can be understood rather easily. Indeed, consider
an imposed current which drives the reacting ions
towards the reaction zone (we shall refer to this current as
{\it forward current}). It is clear that the forward current
enhances the production of $C$-s in the reaction zone. Reversing
the direction of current ({\it backward current}), on the other,
hand works against the reaction and results
in a lower production of $C$-s. Thus, provided the position of
the front is known [i.e. $x_n(t)$ is available from the time-law],
the times $t_n$ of crossing of the threshold concentration and,
consequently, the positions of the band $x_n(t_n)$
can be controlled by an appropriately chosen current.
Since managing the electric current is not an experimental
difficulty, the above method provides us a flexible technique for
the creation of complex precipitation patterns.

\subsection{Mathematical description of the process}
For a more quantitative description of the control-by-current
method, we shall use a mean field model that has been developed
in a series of papers during the last decade
\cite{Pattern-Design,ModelB,Liese-E-front,Liese-E-pattern}.
The first part of the model addresses the irreversible
$A^{+}+B^{-}\to C$ reaction-diffusion process for
totally dissociated electrolytes $A\equiv(A^+,A^-)$
and $B\equiv(B^+,B^-)$ that are initially separated in space.
The evolution equations for the concentration profile of the ions
$a^{\pm}(x,t)$ and $b^{\pm}(x,t)$ are obtained by assuming
electroneutrality on the relevant time and lengthscales
\cite{Liese-E-front} and, for the case of monovalent ions
with equal diffusion coefficients, the equations are as follows \cite{Pattern-Design}
\begin{eqnarray}
\partial_t a^+&=&D\partial^2_x a^+- j(t)\,\partial_x(a^+/\Sigma)-ka^+b^-
\label{e-1}\\
\partial_t b^-&=&D\partial^2_x b^-+ j(t)\,\partial_x(b^-/\Sigma)\,-ka^+b^-
\label{e-2}\\
\partial_t a^-&=&D\partial^2_x a^-+ j(t)\,\partial_x(a^-/\Sigma) \label{e-3}\\
\partial_t b^+&=&D\partial^2_x b^+- j(t)\,\partial_x(b^+/\Sigma)
\label{e-4} \, .
\end{eqnarray}
Here $D$ is the diffusion coefficient of the ions,
$j(t)=I(t)/{\cal A}$ is the externally controlled electric
current-density flowing through the tube of cross section
${\cal A}$, and $\Sigma=q(a^++a^-+b^++b^-)$ with $q$ being
the unit of charge. The reaction rate $k$ is taken to be
large resulting in a reaction zone of negligible width.
Note that this assumption is compatible with the typical
reactions used in experimental setups producing
Liesegang structures.

The second part of the model explains the pattern formation
through the separation of the reaction product $C$
into a high- and low-concentration phases. The evolution of
the concentration $c(x,t)$ is obtained from the Cahn-Hilliard
equation with the addition of a source term corresponding
to the rate of the production of $C$-s ($ka^+ b^-$)
\cite{ModelB,Liese-E-pattern}. The free energy driving the
phase separation is assumed to
have minima at some low ($c_l$) and high ($c_h$)
concentrations of $C$ and, furthermore, it is assumed to have
the Landau-Ginzburg form in the shifted and rescaled
concentration variable $m=(2c-c_h-c_l)/(c_h-c_l)$. In
terms of $m$, the equation describing the phase-separation
dynamics takes the form \cite{ModelB}
\begin{equation}
\partial_t m=-\lambda\Delta( m - m^3 + \sigma \Delta m)+S(x,t)\; .
\label{cahn-hilliard}
\end{equation}
Here $S(x,t)=2ka^+b^-/(c_h-c_l)$ is the source term coming
from equations (\ref{e-1}--\ref{e-4}). The two parameters
$\lambda$ and $\sigma$ are fitting parameter at this stage,
they can be chosen so as to reproduce the correct experimental
time- and lengthscales \cite{RZ2000,Liese-E-pattern}.

Equations (\ref{e-1}-\ref{cahn-hilliard}) are a closed set of
equations for the concentrations of the electrolytes
($a^\pm$, $b^\pm$) and of the reaction product $c$. Together
with the specification of the initial- and the boundary conditions,
they provide the mathematical formulation of the problem.
Below we shall consider numerical solutions of these equations
which were obtained by the classical
fourth-order Runge-Kutta method.

\subsection{Simulation results - properties of the front
in the presence of a current}
In order to obtain a more detailed understanding of the
front dynamics, we studied
the numerical solution of eqs.~(\ref{e-1}) with initial
conditions of separated electrolytes
[$a^\pm(x<0,t=0)=a_0$, $a^\pm(x>0,t=0)=0$,
$b^\pm(x<0,t=0)=0$, $b^\pm(x>0,t=0)=b_0$], and monitored both
the position of the front $x_f(t)$ and the rate of
production $S=kab$ of the $C$-s (a brief account of these
simulations has appeared in \cite{Pattern-Design}).
The physical parameters in the equations
were chosen to be close to the experimentally relevant values
($a_0/b_0=100$, $D=1.22\cdot10^{-9}\;\mbox{m}^2/\mbox{s}$,
$\sigma=10^{-8}\;\mbox{m}^2$, $\lambda=0.17\cdot10^{-9}\;\mbox{m}^2/\mbox{s}$) and we considered following
scenarios for the current. The {\it no current}
case was used to reproduce
the known front properties (i-iii). {\it Constant forward
and backward currents} of amplitude
$\vert j\vert=10\;\mbox{A}/\mbox{m}^2$ were simulated to
check whether the time law holds on the experimentally
relevant timescales. Finally, we studied
the most interesting case of an {\it alternating current} of constant
absolute value ($\vert j\vert=10\;\mbox{A}/\mbox{m}^2$)
with the sign of it changing in a square wave pattern
at times $\tau n^2$, with $n=0, 1, 2,\ldots$, and $\tau$
fixing the timescale of the protocol.
\begin{figure}[htbp]
\begin{center}
\includegraphics[width=16cm, clip]{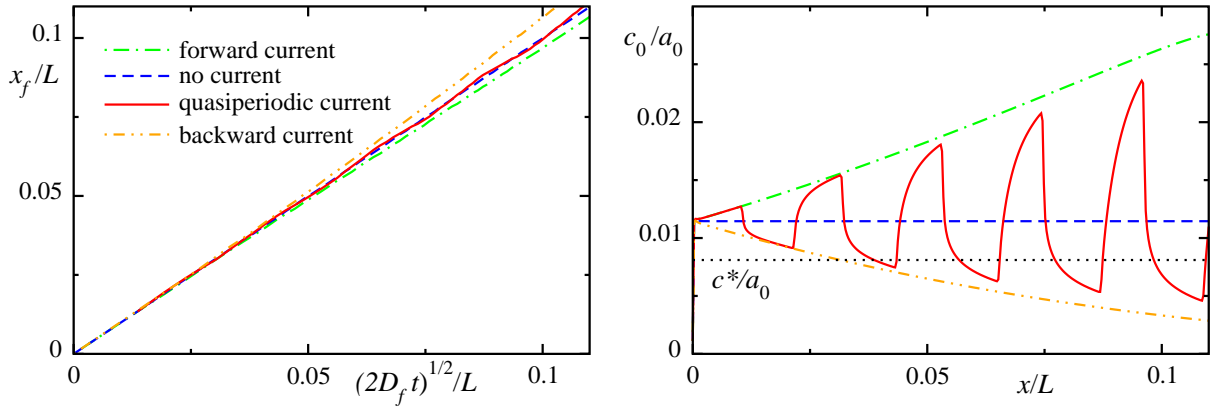}
\caption{Left panel: The position of the reaction front (measured in
units of the length $L$ of the gel column) versus
diffusion length in the absence of a current, displayed
for constant
forward or backward current, and for a quasiperiodic current
(changed at times $\tau n^2$).
Right panel: Concentrations of the reaction product in the
wake of the front for the cases considered on the left panel \cite{Pattern-Design}.
}
\label{fig-front}
\end{center}
\end{figure}

The results are shown on Fig.\ref{fig-front}.
The left panel displays the front motion and shows that in
all cases the diffusive nature of the front is hardly changed,
i.e. the time law (a) remains valid.
On the other hand, as can be seen on the right panel
of Fig.\ref{fig-front},
the $C$-production at the front is strongly influenced
by the character of the applied current. Backward current
leads to a decrease of the concentration of $C$-s left behind
the front, and $c_0(x)$ reaches values below the phase
separation threshold $c_0(x)<c^*$ thus eliminating the possibility
of precipitation. In contrast, the forward current
increases the $C$-production steeply and brings the
system quickly into the unstable regime thus inducing
precipitation. It follows then that, in the case of a
quasiperiodic current, the phase separation can be triggered
and timed by switching on the forward field.

\section{Pattern design}
\label{width}

\subsection{Controlling the spacing and the width of the bands}
Once the front dynamics, summarized in Fig.\ref{fig-front},
is understood, one can invent appropriate current dynamics that
results e.g. in equidistant band patterns \cite{Pattern-Design}.
The wavelength $d$ of the periodic pattern can be predesigned
by switching on the forward currents at
times $t_n=(2 n)^2\tau$, where $n=0,1,2,\ldots$ and
$\tau=d^2/8D_f$. If the desired period $d$ is smaller than
half the {\it local wavelength} of the Liesegang pattern which
would be present without the current (see Fig.\ref{fig-front-and-band}),
then spurious bands may appear
due to a natural increase of the concentration of $C$-s.
This can be avoided, however, by switching on the backward
current when the front is halfway between $x_n$ and $x_{n+1}$,
i.e., at times $(2n+1)^2\tau$.

One can also create more complex patterns both experimentally
and theoretically \cite{Pattern-Design}. When creating patterns 
of several wavelengths, variable widths of the bands may also 
become an important part of the patterns. The issue of the width 
control, treated below, is the novel aspect of the present paper.

Let us begin by finding an estimate of the width of
the equidistant bands having a period $d=\sqrt{8D_f\tau}$.
From simulations we know that even in the presence of
current, the
position of the front is well approximated by
$x_f(t)=\sqrt{2 D_f t}$ where $D_f$ is given by \cite{GR1988}:
\begin{equation}
D_f=2 D \left\{\mbox{erf}^{-1}\left[\frac{a_0/b_0-1}{a_0/b_0+1}\right]\right\}^2\;.
\end{equation}
For a typical ratio of the initial concentrations of A and B, $a_0/b_0=100$, this yields $D_f= 5.43 D$, where $D$ is the diffusion constant of the ions.

\begin{figure}[htbp]
\begin{center}
\includegraphics[width=8cm, clip]{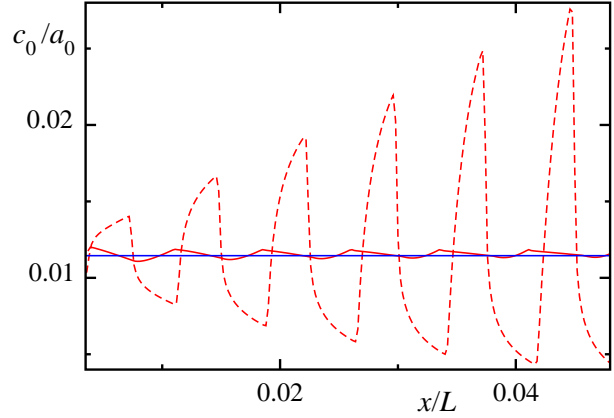}
\caption{$C$-concentration in the wake of the front for the equidistant pattern (red dashed line) and its average over a region of distance $d$ (red solid line) in comparison with the no-current case (blue line).}
\label{fig-c-prod}
\end{center}
\end{figure}

An important point now is that simulations of the equidistant case
[i.e. when the forward currents is switched on at times $t_n=(2 n)^2\tau$]
reveal (see Fig.\ref{fig-c-prod}) that, although the $C$-production
varies strongly within a period, the average concentration
is practically equal to the zero current
case. This means that, for the estimation of the width,
we can replace the complicated function
$c_0(x)$ by the result of the homogeneous production of
$C$-s \cite{MatPack98}:
\begin{eqnarray}
c_0(x)&\approx& c_0=a_0\frac{1+b_0/a_0}{2\sqrt{\pi}}e^{-D_f/(2D)}\sqrt{\frac{2D}{D_f}}\;.
\end{eqnarray}
For $a_0/b_0=100$ this yields $c_0/a_0\approx 1.145\times 10^{-2}$.

Using now the conservation of $C$-s, namely that the amount of $C$-s
produced in a period ($d\,c_0$) is distributed into the high-
and low-concentration regions of length $w$ and $d-w$, the width
of the bands, $w$, is calculated as
\begin{eqnarray}
w&=&\frac{c_0-c_l}{c_h-c_l} d=\frac{c_0-c_l}{c_h-c_l}
\sqrt{8 D_f \tau}\;,
\end{eqnarray}
and we find that the width depends on $\tau$ in the same
way as the period. The width also depends on the diffusion
constant of the front, $D_f$ which, in turn, depends on the
ratios of initial concentrations $a_0/b_0$ as well as on
the diffusion coefficients of the electrolytes.
Most easily, however, the width can tuned by the initial
concentration of $A$, $a_0$ which is proportional to $c_0$.

\begin{figure}[htbp]
\begin{center}
\includegraphics[width=16cm, clip]{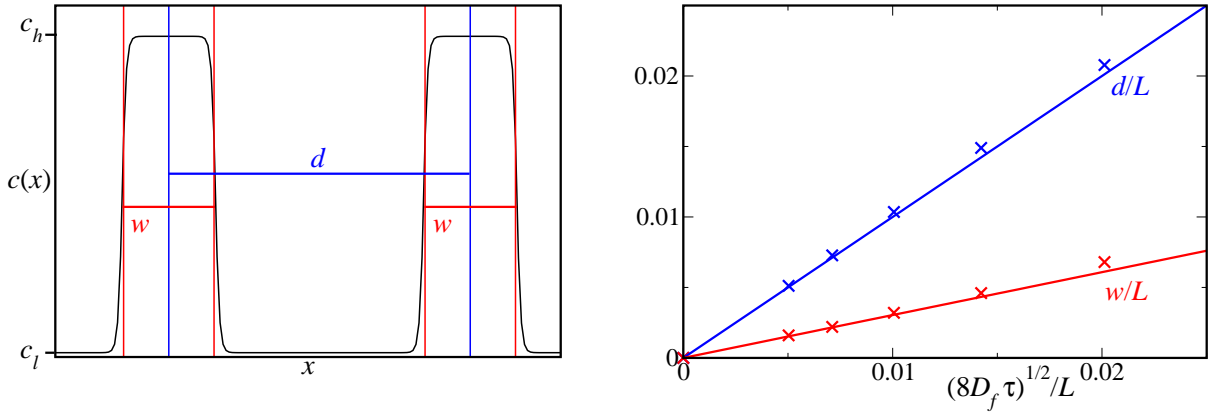}
\caption{Left panel: Definition of the width $w$ and the period $d$
of the equidistant pattern. Right panel: Dependence of
the numerically obtained widths (red cross) and periods
(blue cross) on $\tau$ in comparison with the theoretical
prediction (solid lines).}
\label{fig-bands}
\end{center}
\end{figure}

In Fig.\ref{fig-bands} one can see the numerical verification
of the above approximations for the width and the wavelength
of the periodic pattern. The smaller the value of $\tau$
the better is the estimate, since for large values of
$\tau$ we enter a regime where the deviations from the
time-law (a) become significant.

\subsection{Encoding information into precipitation structures}

\begin{figure}[htbp]
\centering\includegraphics[width=12cm, clip]{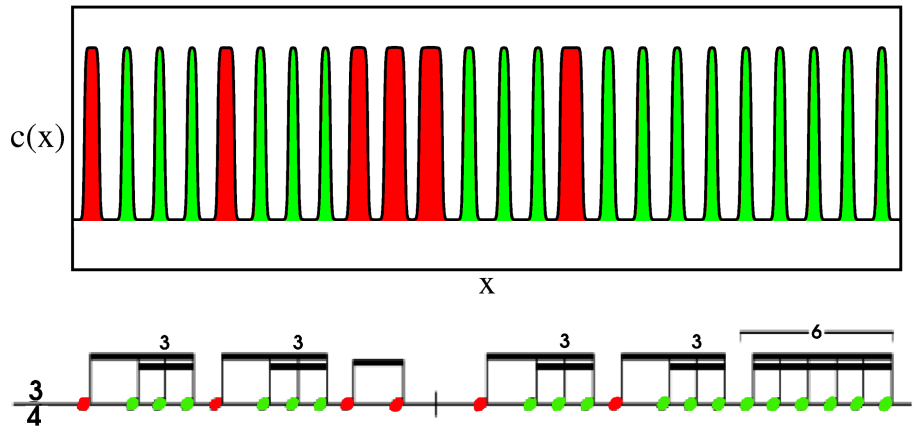}
\caption{Musical rhythm written into Liesegang bands.
The upper panel shows the concentration-profile of narrow and wide bands
(the first four bands are cut out of the picture as well as the
final ones since they don't belong to the implementation
of the rhythm). This is compared in the lower panel to
the Bolero rhythm composed by Ravel \cite{footnote}
.}
\label{fig-bolero}
\end{figure}

\begin{figure}[htbp]
\centering\includegraphics[width=12cm, clip]{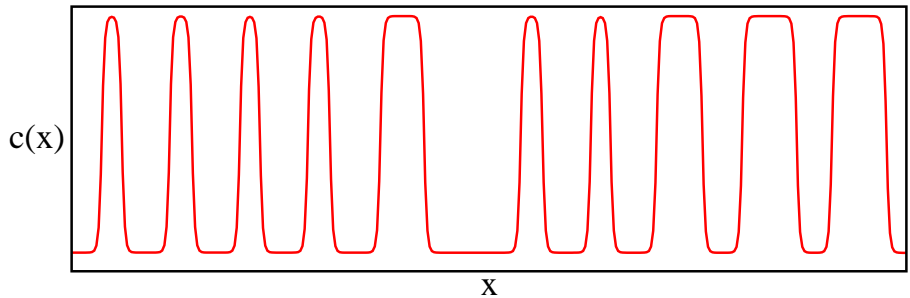}
\caption{Concentration profile, $c(x)$ for a numerical
test of the combined width and space control.
In the simulation the number ``42'' is written into
precipitation bands based on the Morse alphabet:
4 (....\_) 2 (..\_\_\_). The dots are implemented
by narrow bands, the underscore by a wide band approximately 
a factor $f=2$ larger, and the characters are separated
by spaces, widened by the same factor $f$.}
\label{fig-42}
\end{figure}

Since spacing and width can be controlled by properly designed current,
more complex structures should be realizable. The pattern shown in Fig.~\ref{fig-bolero} has been created using a slightly
generalized method of control described above for the periodic pattern.
The underlying pattern in Fig.~\ref{fig-bolero} is an equidistant
pattern but the width of certain bands is increased.
This has been achieved by a longer time interval for the forward
field for the wide bands. More precisely, in order to make
the width of the $n$-th band larger, the switching protocol
is changed: instead of switching on the backward current at time $(2n+1)^2\tau$, it is switched on at time
$((2n+f)^2\tau$ such that the band appears approximately a factor $f$ larger. Similarly it is also possible to increase the spaces between the bands by increasing the time duration of the backward current. In Fig.~\ref{fig-42} we demonstrate this by creating a structure
(the Morse code of the famous number 42 \cite{adams}) where a 
current dynamics had to be designed which yields simultaneous 
control of the spacing and the width of the bands.

\section{Conclusions and outlook}
We have developed further the techniques
of the engineering of precipitation structures
using electric currents.
It was shown that an appropriately designed time-dependent
current allows to control not only the band-spacing but also
the width of the bands in one-dimensional structures. This
permits the encoding of information by controlling either only the
the width (as exemplified by the Bolero rhythm) or by combining the
width and space control (as was shown on the example of the
famous number 42).

A naturally arising question concerns the limits of the approach
when trying to downscale the structures. As far as the timescales
are considered, since the diffusion coefficients of the
participating agents are of order $10^{-9}$ $\mbox{m}^2$/s,
imposed currents on the time scale of
$\tau =0.1 s$ provide control on
lengthscales of $10\mu$. Thus scaling the dynamics of the current
does not pose an experimental problem.

The real problem with
downscaling is the width of the bands. The minimum width
is clearly related to the width
of the reaction front which is not negligible on
the scale of microns and, depending on the reagents, can be even
at the scale of $mm$-s \cite{Koo1991}. It should be noted however that 
spontaneous pattern formation on the nanoscale level has 
been observed \cite{Mohr}, and these type of systems
might be appropriate for the pattern control by imposed 
electric currents.
Another problem with the width of the bands is that the thermal fluctuations 
in the concentrations of the reagents combined with those of the gel make the
reaction front uneven and, depending on the surface tension
of the created bands, they may lead to a roughening of
the band on a scale that is comparable to the width.
Clearly, advances in downscaling can be achieved only
if models are developed which can treat all the
above problems.

Finally, we note that combining the proposed technique
with the already existing indirect control  strategies
such as the choice of geometry and initial conditions opens
up a wide spectrum of possible structure design.
So far the feasibility of the approach has
only been shown for a few types of predesigned
one-dimensional patterns. A straightforward extention
would be the creation of rings or spheres with a
predesigned internal structure. Further, the extention of the
current-control technique to stamping methods \cite{grzybowski}
on the mesoscopic level could be used to create even
more complex pattern designs which may become useful in
engineering applications.

\ack
This work has been partly supported by the Swiss National Science
Foundation and  by the Hungarian Academy of Sciences (Grant
No.\ OTKA K68109).

\end{document}